# Chiral nanoparticle chains on inorganic nanotube templates

*Lukáš Kachtík[1], Daniel Citterberg[1], Kristýna Bukvišová[1], Lukáš Kejík[1], Filip Ligmajer[1,2], Martin Kovařík[1,2], Tomáš Musálek[1,2], Manjunath Krishnappa[3], Tomáš Šikola[1,2] and Miroslav Kolíbal[1,2]\**

[1]CEITEC BUT, Brno University of Technology, Purkyňova 123, 612 00 Brno, Czech Republic

[2]Institute of Physical Engineering, Brno University of Technology, Technická 2, 616 69 Brno, Czech Republic

[3]Faculty of Sciences, Holon Institute of Technology, 52 Golomb St., Holon, 5810201, Israel

*kolibal.m@fme.vutbr.cz

**Abstract**

Fabrication of chiral assemblies of plasmonic nanoparticles is a highly attractive and challenging task with promising applications in light emission, detection, and sensing. So far, primarily organic chiral templates have been used for chirality inscription. However, this significantly limits the variety of nanoparticle preparation techniques to an in-pot approach at very low synthesis temperatures. Here, we demonstrate utilization of seemingly achiral inorganic nanotubes as templates for the chiral assembly of nanoparticles. We show that both metallic and dielectric nanoparticles can be attached to scroll-like chiral edges propagating on the surfaces of $WS_2$ nanotubes. Due to relatively high temperature stability of these nanotubes, such assembly can be performed at temperatures as high as 550 °C. This large temperature range significantly widens the portfolio of usable nanoparticle fabrication techniques, allowing us to demonstrate a variety of chiral nanoparticle assemblies, ranging from metals (Au, Ga) and semiconductors (Ge) to oxides ($WO_3$).

**Manuscript text**

Many important biomolecules and pharmaceuticals are chiral because of the inherent chirality of life at our planet[1]. The interaction of these chiral molecules with light can be used for discriminating between them, controlling them, or even synthesizing them. The large discrepancy between the size of the molecules and the wavelength of light, however, severely hinders these promising applications. It is therefore desirable to prepare chiral nanostructures and metamaterials that exhibit exceptionally strong light-matter interactions, potentially increasing the sensitivity and throughput wherever the chirality light-matter interactions come into question.

The strength of chiroptical effects can be characterized by the dissymmetry factor (g-factor) defined as a ratio of differential extinction of circularly polarized light to total extinction. Very large g-factors can be reached by squeezing the light waves down to the length scale of biomolecules by carefully designed plasmonic[2,3] or dielectric[4,5] nanostructures. Their chiroptical response can be further enhanced by inscribing a long-range chiral order to extended clusters of nanoparticles.[6–8] Such rationally designed assemblies can be tailored to exhibit g-factors in the visible part of the electromagnetic spectrum, enhanced by higher-order collective interactions.[9,10] These chiral superstructures are an attractive candidate for plasmon-enhanced chiral sensing where the single-molecule sensing potential of plasmonics[11] is utilized in the realm of chirality analysis.[12] Although there is an ongoing debate whether achiral sensing platforms could be superior to the chiral ones,[13] the sensing capability of chiral superstructures has already been demonstrated in the detection of markers of Parkinson's disease[14] or in the detection of attomolar DNA concentrations.[15] Going beyond sensing, chiral nanoparticle assemblies can be also used for driving chiral photochemistry and facilitating enantioselective chemical synthesis,[16,17] which could have a disrupting effect on production of pharmaceuticals. Chiral nanoparticle assemblies could also serve as sources or detectors of circularly polarized light,[18,19] dynamic chiral nanomachines,[20] or in chiral phototherapy.[21]

The strategies to prepare chiral nanoparticle assemblies comprise two approaches: top-down and bottom-up. The former, usually lithography-based, yield complex geometries[22,23] but are time-demanding and it is difficult to scale them up. The latter, on the other hand, utilize either some sort of symmetry breaking during a nanoparticle synthesis process[24–27] or chirality transfer from an already existing chiral template to achiral nanoparticles.[19,28–30] The vast majority of the chiral templates are 'soft' like, e.g., DNA molecules,[9] amino acids,[31,32] micelles,[33] chiral polymers,[34,35] chiral ligands in perovskites,[36] peptides,[37] or proteins.[14] Unfortunately, these 'soft' templates are fragile and can easily be destroyed by elevated temperatures required in the subsequent synthesis and processing steps. Moreover, the bonding flexibility within the soft template can result in changes in interparticle distances (e.g. in a liquid environment) and even disorder of the original helical arrangement.[37]

Utilizing a 'hard' inorganic template would naturally offer more degrees of freedom for its further processing steps or functionalization with optically active nanoparticles. However, such synthesis has been demonstrated only using silica nanohelices so far, providing a template for gold[17,38] or perovskite nanocrystal assemblies.[39] Moreover, these studies demonstrated only a wet chemistry-based approach, which is limited to in-pot chemical reactions and their products. Going beyond this approach, here we utilize chiral grain boundaries present on the surface of inorganic nanotubes made of a layered van der Waals material ($WS_2$) that serve not only as attachment and nucleation sites for nanoparticles prepared by wet chemistry, but also during evaporation under vacuum and high-temperature conditions. The resilience of inorganic nanotubes with respect to high temperatures opens the possibility of bottom-up fabrication of chiral assemblies of nanoparticles made of a much broader group of materials, including those not available so far, like germanium, gallium, or tungsten trioxide.

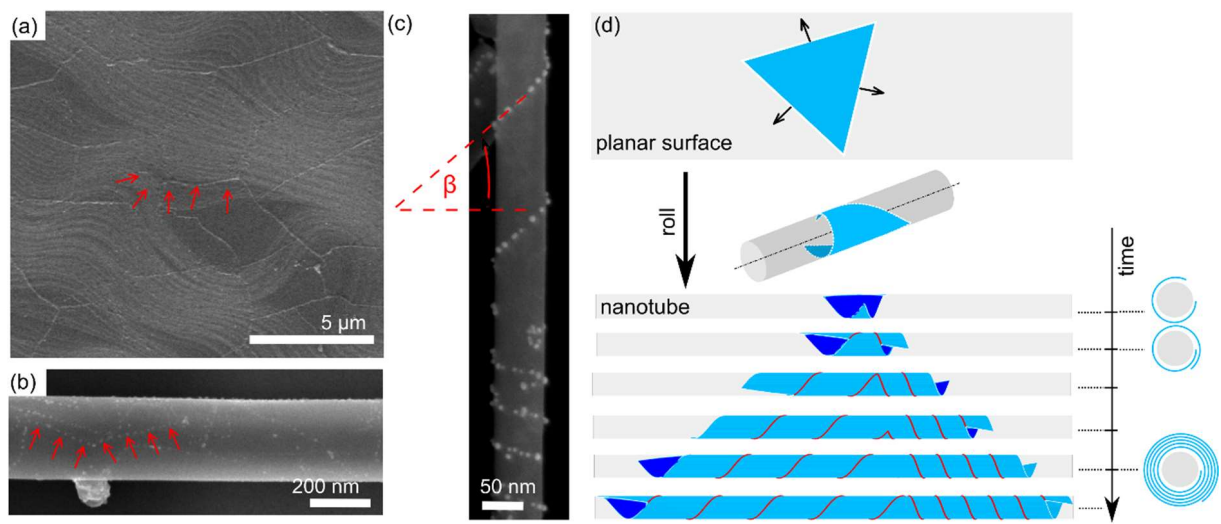

Fig. 1: *Formation of line defects on tubular templates. (a) Mild oxidation of a flat copper foil covered with polycrystalline graphene reveals graphene's grain boundaries via formation of copper oxide beneath, as indicated by the red arrows in the scanning electron microscopy (SEM) image. (b) A similar effect is observed on the surface of a $WS_2$ nanotube, where the grain boundaries have been accentuated by controlled surface oxidation.[40] (c) SEM micrograph of a $WS_2$ nanotube decorated with a helical array of gold nanoparticles. The helix can be characterized by a chiral angle β, as marked in the image. (d) A theoretical model of lateral growth of a triangular domain on an unwrapped plain surface (top) and on a tubular template (bottom). If the two domain edges meet during growth, one of them overgrows the other, eventually resulting in a scroll-like overlayer. Cross sections at different time intervals are shown as well. See Supporting Information for more details of the model and also for yet another scenario, where a grain boundary is formed instead of an edge.*

Line defects (grain boundaries or step edges) on the surfaces of solid-state materials exhibit higher reactivity than the defect-free areas,[41,42] and are thus more vulnerable to chemical attacks such as oxidation or reduction. For example, copper oxide begins to form on a graphene-covered copper foil just beneath the grain boundaries of the polycrystalline graphene, serving as a tool to visualize these otherwise hardly visible boundaries[43] (see Fig. 1a). Notably, crystallographic defects can also be visualized in a similar manner on inorganic nanotubes made of $WS_2$[40] (see Fig. 1b). As the smooth surface of nanotubes is mostly inert,[44,45] the line defects (unsaturated sulfur bonds in the case of $WS_2$) are then the only chemically active sites.[46,47] Due to the tubular shape of the surface, a single-line defect could in principle form a chemically reactive three-dimensional spiral (called a chiral line in the following text), which allows for selective attachment of various molecules or nanoparticles into a helical assembly. To verify this hypothesis, we have decorated a large batch of $WS_2$ nanotubes with gold nanoparticles using an established one-pot chemistry protocol (see Methods for details). During the subsequent inspection, we have found that some of the nanotubes indeed exhibit such helical nanoparticle chains (Fig. 1c).

In order to understand the nature of the line defect, we have modeled two plausible scenarios of the chiral line formation during $WS_2$ formation in a growth reactor. Both scenarios describe the same process: sulfurization of tungsten oxide nanowires in $H_2S$ gas during the formation of $WS_2$ nanotubes in a growth reactor.[48–51] We utilize an analytical model of domain spreading in a tubular geometry (see the Supporting Information for details). For simplicity, we have assumed that growth of an initial $WS_2$ domain on the surface of WOx nanowire could occur from only a single nucleation site (see Supporting Information, Fig. S1). We assume this initial domain to be of triangular shape (Fig. 1c) based on the frequent observations in tungsten oxide sulphurization experiments.[52] The lateral growth of the triangular domain on a cylindrical template inevitably includes a collision of the growing edge with another edge of the very same domain. In the first scenario, the collision leads to formation of a domain boundary, finally resulting in a single helical line defect within an outermost single layer of a nanotube (Fig. S1a). This line defect is characterized by a chiral angle $\beta$ (see Fig. 1c). The final morphology of a decorated nanotube exhibits only a limited range of the chiral angles (see Supporting Information, Fig. S1b). We will show later that this scenario does not predict all the experimental results. Nevertheless, it could potentially reproduce line defect formation under different nanotube growth conditions or in different material systems. The second scenario corresponds to overgrowth of an already formed nanotube by an additional $WS_2$ layer, e.g. from a supersaturated vapor during the growth-reactor cooling down. Fig. 1d shows the evolution of such an overlayer's growth. We have again assumed the initial domain to be triangular, which is a common shape resulting from chemical vapor deposition of transition metal dichalcogenides (TMDs) in general and $WS_2$ in particular.[53] In contrast to the first scenario, the propagating layer was allowed to climb over the other edge upon their collision and continue to grow laterally as an overlayer (Fig. 1d). The resulting step edge then forms the chiral helical line. The final morphology in this scenario is a scroll-like multiwall nanotube (Fig. 1d, bottom) instead of an in-plane grain boundary. Notably, there are always two chiral lines corresponding to the two sides of the initial triangular domain. The chiral angles of the two lines are naturally affected by the orientation of the initial domain with respect to the nanotube's primary axis. This orientation is not restricted by any geometrical or physical constraints, and hence neither are the chiral angles. Furthermore, the geometry of the initial triangular domain dictates that the angle between the two chiral lines is always 60°. These anticipated characteristics match well with the results of nanotube decoration experiments (see Supporting Information, Fig. S3). Additionally, a transmission electron microscopy (TEM) analysis of the nanotubes decorated with nanoparticle helices confirmed the scroll-like nature of their outer walls (see an example in Fig. 2).

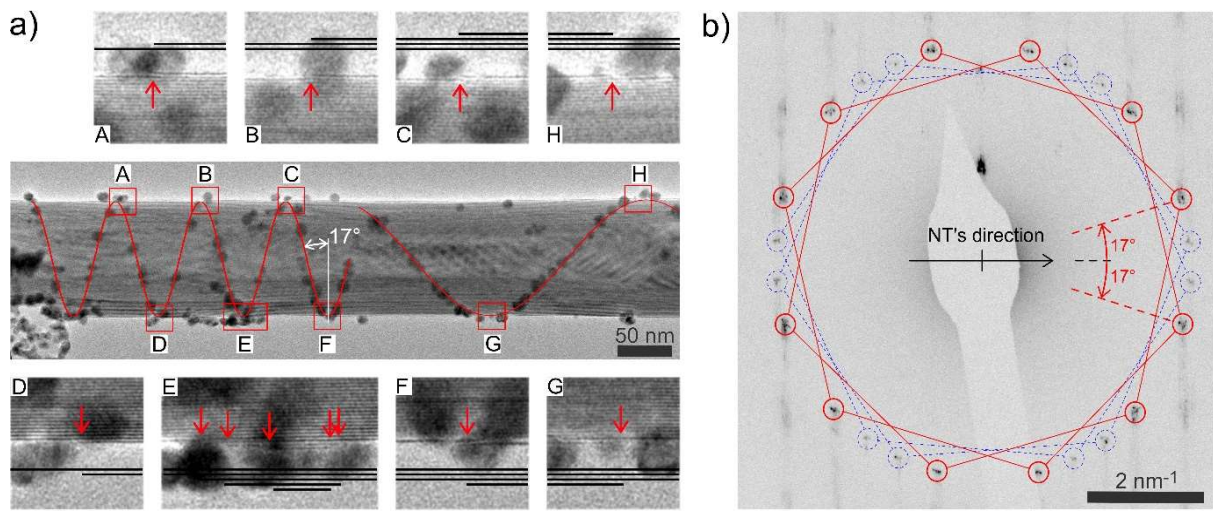

Fig. 2: *TEM analysis of a WS$_2$ nanotube decorated with gold nanoparticles. (a) The TEM image in the center shows a multiwalled WS$_2$ nanotube decorated with gold nanoparticles along a chiral line highlighted by the two red sine-shaped curves. Each turning point of the chiral line at the nanotube edge is shown in detail (A-H close-ups). In the close-ups, the individual WS$_2$ walls are visible as dark horizontal lines and are highlighted using schematic black lines in sketches above or bellow the original images. The close-ups confirm that the chiral line is, in fact, a step edge (marked by red arrows) of an outer WS$_2$ wall. The close-ups are 25 nm in height. (b) TEM diffraction pattern taken within the A-F region of the same nanotube shown in (a). The pattern comprises two chiral angles (blue and red). One*

*of them (17°) is identical to the one measured from the real space image of the nanoparticle chain (chiral line). The other one (blue) corresponds to the inner wall(s) of the multiwall nanotube.*

Figure 2a shows a TEM image of a $WS_2$ nanotube decorated with a spiral chain of Au nanoparticles. A detailed inspection of the image reveals noticeable steps at the nanotube's surface, located precisely at the positions of the chain's turning points (Fig. 2a, close-ups). The step sequence, taken from left to right, consistently exhibits a step-up until a point marked F is reached. From then on, only step-downs are present. Such a sequence indicates that the outer envelope of the nanotube is formed by a rolled wall, which creates a scroll-like tubular surface wich a helical chiral line. Every nanotube with a chiral nanoparticle chain we inspected in TEM exposed such a scroll-like overlayer. Such an observation is consistent with the second scenario discussed above (Figure 1d). Figure 2b shows the TEM diffraction pattern of the very same nanotube. The diffraction spots can be grouped into two pairs of colored hexagons, where each pair corresponds to a single chiral angle (because the electron beam passes both the top and the bottom part of the nanotube). The existence of multiple hexagons is due to the multi-wall nature of the nanotubes, where each inner wall is randomly oriented against each other and can be assigned a chiral angle. However, one of them (17°, red color in Fig. 2b) is always identical to the one measured in the real space image (Fig. 2a). We have similarly analyzed several other nanotubes (see Tab. S1, Supporting information) and were always able to find a match between the chiral angles measured from the respective diffraction pattern and those measured directly from an SEM image of the nanoparticle chains, thus generalizing our conclusions. A thorough TEM analysis allows us to go even beyond just chiral angle determination: it is possible to identify the nature of step edges (see Supporting Information, Fig. S4 and Table S1). We have identified them to be of zigzag type, which is consistent with predictive models using density functional theory.[52] Having identified the chiral lines as step edges instead of grain boundaries makes possible to utilize the step edge as a nucleation site for adatoms diffusing along the nanotube surface, e.g. during vapor deposition. We demonstrate this capability in last part of the paper.

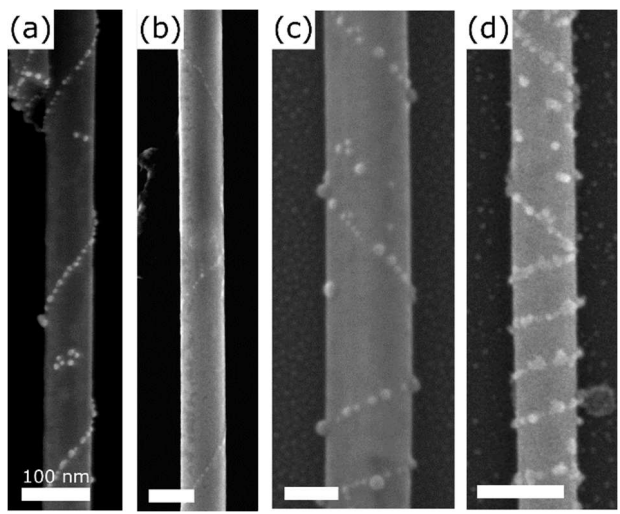

*Fig. 3: Decoration of chiral WS$_2$ nanotubes by various techniques and materials (see Methods for details). SEM images of: (a) Gold-decorated nanotube, resulting from reduction of elemental gold out of an aqueous solution of HAuCl$_4$. (b) Nanotube decorated with WO$_3$, resulting from oxidation by water vapor at 300 °C. (c) Gallium- and (d) germanium-decorated nanotubes resulting from thermal evaporation of these elemental materials in a vacuum onto dry WS$_2$ nanotubes pre-deposited by drop-casting onto a solid substrate (silicon).*

Gold nanoparticle assemblies similar to those studied above have been previously demonstrated (albeit with different geometry and dimensions).[9,38] However, the advantage of our inorganic tubular templates over 'soft' organic templates is the possibility of utilizing higher process temperatures, therefore, wider variety of materials for nanoparticle assemblies can be used. This is demonstrated in Fig. 3b, where a WS$_2$ nanotube has been controllably oxidized in water vapor at 300 °C,[40] leading to the formation of WO$_3$ nanoparticles along the edge of the outer wall. The very high temperature stability of WS$_2$ nanotubes (up to at least 550 °C in high vacuum, as confirmed by a correlative SEM and XPS study)[40] also offers flexibility in terms of deposition techniques. Figs. 3c,d show WS$_2$ nanotubes decorated with gallium and germanium nanoparticles, which were deposited by evaporation of the respective elemental materials under high-vacuum conditions onto nanotubes dispersed over a solid substrate (silicon). Note that gallium could be subsequently oxidized to form gallium oxide or nitrided to form gallium nitride,[54] thus widening the portfolio of available nanoparticles. There are no limitations other than the deposition temperature that would prevent other techniques like atomic layer deposition[55] or chemical vapor deposition[56] from being used for creation of chiral nanoparticle assemblies utilizing these inorganic templates.

Note that the scroll-like WS$_2$ nanotubes discussed here are rarely found within bare unsorted nanotube batches because the synthesis processes for multi-wall WS$_2$ nanotubes with smooth outer surface are currently well developed.[50] Here, we have shown that the initially undesired scroll-like products of nanotube synthesis can become templates for for further processing towards very attractive nanostructures. Hence, further effort is necessary to identify the growth conditions resulting in the formation of scroll-like nanotubes or to invent some efficient sorting techniques . We hypothesize that the scroll-like overlayer forms when a reactor slowly cools down from growth temperature during the preparation of WS$_2$ nanotube templates. At lower temperatures, it is plausible that the vaporized transient tungsten oxides resulting from the nanotubes' formation process condense on the outer walls of a few nanotubes and react with the H$_2$S residues in the reactor. This CVD-like reaction results in WS$_2$ nucleation and overlay growth on the surface of existing WS$_2$ nanotubes, following the scenario depicted in Fig. 1d. Modification of the reactor conditions during cooling could be a possible way toward the formation of scroll-like nanotubes with high yield. Furthermore, tubular inorganic templates must not be limited only to nanotubes: Nanoscrolls formed by rolling up 2D materials have been experimentally observed since 2003.[57–59] E.g., carbon nanoscrolls can withstand even higher temperatures that TMDs and, thus, could serve as templates for formation of nanoparticle chains that require even higher synthesis temperatures than shown in this study. Nevertheless, despite recent attempts to clarify the rolling process of planar 2D layers, the overall understanding of the formation of the nanoscrolls is still poor.[60,61] With further advancements in the preparation of these nanostructures, the utilization of the (chiral) defect lines for nanoparticle attachment represents a promising way toward (helical) nanoparticle assemblies of various materials and geometries.

In summary, we have demonstrated the fabrication of metallic, semiconductor, and oxide nanoparticle chiral assemblies by utilizing scroll-like inorganic nanotube templates. In contrast to the commonly used organic templates, WS$_2$ nanotubes are stable up to 550 °C in vacuum. Taking advantage of the very high temperature stability of these templates, we have demonstrated the possibility of preparing a wide range of nanoparticle assemblies (metals, oxides, semiconductors). Using physical vapor deposition allowed going beyond in-pot chemical synthesis which has been exclusively utilized so far in connection with organic chiral templates.[46,47,62] The versatility of our approach holds promise for fabrication of chiroptically active building blocks in emerging areas of chiral nanophotonics, where facile fabrication of such assemblies represents a great challenge. For example, in magnetochiral dichroism, helical assemblies of magnetic nanoparticles are studied for their higher-order optical response.[63] Similarly, understanding of semiconductor chiroptical nanomaterials also remains elusive, despite their attractiveness in many fields.[64] Besides, complexes of TMD-templates with plasmonic nanoparticles could become an ideal playground for studying chiral plasmon-exciton polariton complexes.[65] The proposed fabrication approach thus significantly advances the so-far narrow range of preparation techniques of chiral nanomaterials, allowing for experimental studies of the theoretically proposed complexes.

**Methods**

**Synthesis of $WS_2$ inorganic nanotubes**

The $WS_2$ nanotubes were prepared via 'vapor-gas-solid (VGS)' method, as discussed in detail in our previous reports.[50,51] In a brief, the $WO_3$ precursor was initially reduced to tungsten suboxide ($WO_{2.75}$) phase using $H_2$ gas as a reducing agent, which results in formation of 1D $WO_{2.72}$ suboxide nanowhiskers. The as-formed $WO_{2.72}$ nanowhiskers undergo further sulfurization as a 'solid−gas' reaction under the continuous flows of $H_2S$ and $H_2$ gases to convert suboxides into multi-walled, hollow tungsten disulfide nanotubes.

**$WS_2$ NT-AuNP complex preparation**

We decorated $WS_2$ nanotubes with gold nanoparticles using a chemical procedure described in Ref. 46. Briefly, 2.6 mg of powdered $WS_2$ nanotubes were dispersed in 2 ml of isopropyl alcohol (IPA) using a combination of an ultrasonic bath and subsequent mechanical shaking; each step was repeated at least four times. The WS$_2$ nanotube dispersion was then added to a hot plate-heated beaker with boiling 0.043 mM aqueous solution of HAuCl4 under vigorous stirring. The dispersion was still being heated to boil for an additional 3 min, and subsequently the hot plate was turned off and let cool down slowly to a room temperature. This procedure resulted in WS$_2$ nanotubes decorated with gold nanoparticles of an average diameter around 10 nm. By varying the respective ratio between WS$_2$ and HAuCl$_4$, e.g. by changing the HAuCl$_4$ concentration, the average diameter of the AuNP could be tuned between 5 and 30 nm. The resulting dispersion was then drop-casted onto a cleaned substrate (silicon wafer, glass covered with an indium tin oxide film, or TiN membrane).

**Evaporation under vacuum conditions**

As described above, nanotubes were dispersed in IPA using an ultrasonic bath and mechanical shaking. Next, the solution was drop-casted onto a silicon substrate and let dry. The resulting sample was inserted into an ultrahigh vacuum chamber (base pressure $p = 1.10^{-10}$ mbar) equipped with gallium and germanium evaporators. The sample was first annealed using a calibrated pBN heating element at 350 °C for 20 minutes to degas any IPA residues and then stabilized at growth temperature (215 °C and 360 °C for gallium and germanium, respectively). The growth temperature, flux, and overall duration were optimized to achieve nanoparticle formation on the chiral line only. For gallium, the optimum flux was 1.08 nm/min for 2 minutes. For germanium, the optimum flux was 0.05 nm/min for 10 minutes.

**Acknowledgments**

We acknowledge the institutional support of FME BUT (grant no. FSI-S-20-6485). M. Kov. Was supported by specific research grant CEITEC VUT-J-21-7372. CzechNanoLab project LM2018110 funded by MEYS CR is gratefully acknowledged for the financial support of the sample fabrication at CEITEC Nano Research Infrastructure.